\documentclass[conference]{IEEEtran}
\IEEEoverridecommandlockouts
\usepackage{graphicx}
\usepackage{epsfig}
\usepackage{latexsym}
\usepackage{amsfonts}
\usepackage{here}
\usepackage{rawfonts}
\usepackage[latin1]{inputenc}
\usepackage[T1]{fontenc}
\usepackage{calc}
\usepackage{url}
\usepackage{enumerate}
\usepackage{color}
\usepackage[tbtags]{amsmath}
\usepackage{amssymb}
\usepackage{upref}
\usepackage{epic,eepic}
\usepackage{times}
\usepackage{dsfont}
\usepackage{comment}
\usepackage{cite}

\usepackage{graphicx}
\usepackage[font={small}]{caption}
\usepackage[font={small}]{subcaption}

\usepackage{stfloats}
\usepackage{mathtools}
\usepackage{amsmath}

\usepackage{algorithm}
\usepackage{algpseudocode}

\def\BibTeX{{\rm B\kern-.05em{\sc i\kern-.025em b}\kern-.08em
    T\kern-.1667em\lower.7ex\hbox{E}\kern-.125emX}}

\newtheorem{theorem}{\bf{Theorem}}

\newtheorem{remark}{Remark}

\begin{document}

\title{Hierarchical Over-the-Air Federated Learning with Awareness of Interference and Data Heterogeneity}

\author{
	\IEEEauthorblockN{Seyed~Mohammad~Azimi-Abarghouyi and Viktoria~Fodor}
	\\ \IEEEauthorblockA{KTH Royal Institute of Technology and Digital Futures, Stockholm, Sweden, \{seyaa,vjfodor\}@kth.se}
	\thanks{An extended version of this work is under review in \cite{journal}.}
	\vspace{-10pt}
}

\maketitle

\begin{abstract}
	 When implementing hierarchical federated learning over wireless networks, scalability assurance and the ability to handle both interference and device data heterogeneity are crucial. This work introduces a learning method designed to address these challenges, along with
	 a scalable transmission scheme that efficiently uses a single wireless resource
	 through over-the-air computation. To provide resistance against data heterogeneity, we employ gradient aggregations. Meanwhile, the impact of interference is minimized through optimized receiver normalizing factors. For this, we model a multi-cluster wireless network using stochastic geometry, and characterize the mean squared error of the aggregation estimations as a function of the network parameters.
	We show that despite the interference and the data heterogeneity, the proposed scheme achieves high learning accuracy and can significantly outperform the conventional hierarchical algorithm.
	
\end{abstract}
\vspace{0pt}
\vspace{-5pt}
\section{Introduction}
Machine learning by transferring large amounts of data from edge devices to a central server is rarely feasible due to strict constraints on latency, power, and bandwidth, or concerns on data privacy \cite{viktoria}. A practically feasible distributed approach is federated learning (FL), which implements machine learning directly at the wireless edge, ensuring that the data never leaves the devices \cite{mcmahan}. This approach includes local model training at the devices and model aggregation at a server. To support a high number of collaborating devices and to speed up the learning process, recent studies propose hierarchical architectures with two levels of aggregation with a core server and multiple edge servers \cite{ letaief, bennis, tony, gunduz_ota}. In this paper, we focus on the interference and data heterogeneity that arise when hierarchical FL is implemented in wireless networks, and propose and analyze a learning method and transmission scheme that are tailored to address these specific challenges.

The convergence properties of hierarchical FL, without considering the limitations of a wireless environment is evaluated in \cite{letaief}, and FL with orthogonal transmission is considered for example in \cite{bennis, tony}. Non-orthogonal over-the-air FL has been proposed to avoid the communication bottleneck when a high number of devices participate in the learning. This approach leverages interference caused by simultaneous multi-access transmissions from edge devices to perform model aggregation.  
The majority of research in the field of over-the-air FL focuses on single-cell learning scenarios, with an emphasis on the uplink transmissions \cite{ huang_analog}, and recently also uplink interference. Interferers distributed according to Poisson point processes (PPPs) are considered in \cite{huang_sg}, while an abstract interference model with heavy tail is considered in \cite{yang3}. Studies have been also conducted to examine bandwidth-limited downlink in single- \cite{downlink} and multi-cell \cite{multicell} settings. Hierarchical FL using over-the-air computation is studied in \cite{gunduz_ota}. However, that study assumes the presence of multiple antennas at the edge servers, an ideal downlink transmission, and most importantly, does not account for inter-cell interference. The work we present here is a necessary next step to accurately consider the unavoidable interference. 

To model the effect of interference in hierarchical FL, we can turn to stochastic geometry \cite{haenggi_book}. Within the field, there are two \emph{canonical} approaches to characterize wireless networks. One of them is PPPs, which assumes uniform device and server placements, and has applications in cellular networks \cite{huang_sg}. The other approach is based on Poisson cluster processes (PCPs). PCPs are justified by Third Generation Partnership Project (3GPP) \cite{dhillon} and widely-adopted in information-centric deployments where devices are frequently grouped together in specific areas, known as ``hotspots'' \cite{dhillon, hole}. As the latter fits to the application area of large scale distributed learning, this is the direction we follow in this paper.

This paper develops a learning and transmission scheme for hierarchical FL utilizing over-the-air computation, and provides modeling solution that can capture the effects of interference and data heterogeneity and include them in the system optimization. The key contributions are as follows:

\textit{Learning Method:} We propose a new hierarchical learning method, {\fontfamily{lmtt}\selectfont MultiAirFed}, well suited for unreliable wireless links and non-i.i.d. data. It combines intra-cluster gradient and inter-cluster model parameter based aggregations, as well as multi-step local training.

\textit{Transmission Scheme:} We propose a scalable clustered over-the-air aggregation for the uplink and a bandwidth-limited analog broadcast for the downlink, enabling each iteration all over the network within a single resource block. 
While the proposed transmission scheme is general, we apply it for {\fontfamily{lmtt}\selectfont MultiAirFed}, and express the intra- and inter-cluster aggregation errors caused by the uplink and downlink interference. 

\textit{Tractable Modeling and Optimization:} By using stochastic geometry tools, particularly PCPs, we characterize the distortion of the intra- and inter-cluster aggregations in terms of the mean squared error (MSE). Then accordingly, we optimize normalizing factors for the proposed aggregation estimations.

\textit{Design Insights:} Our analysis indicates that optimized normalizing factors decrease the impact of cluster density on the distortion. Also, simulation results demonstrate that {\fontfamily{lmtt}\selectfont MultiAirFed} notably outperforms the conventional hierarchical FL algorithm, particularly in high-interference and data-heterogeneity scenarios.

\vspace{-4pt}
\section{System Model}
\begin{figure}[tb!]
	\vspace{-123pt}
	\centering
	\hspace*{-.7cm}  
	\includegraphics[width =4.5in]{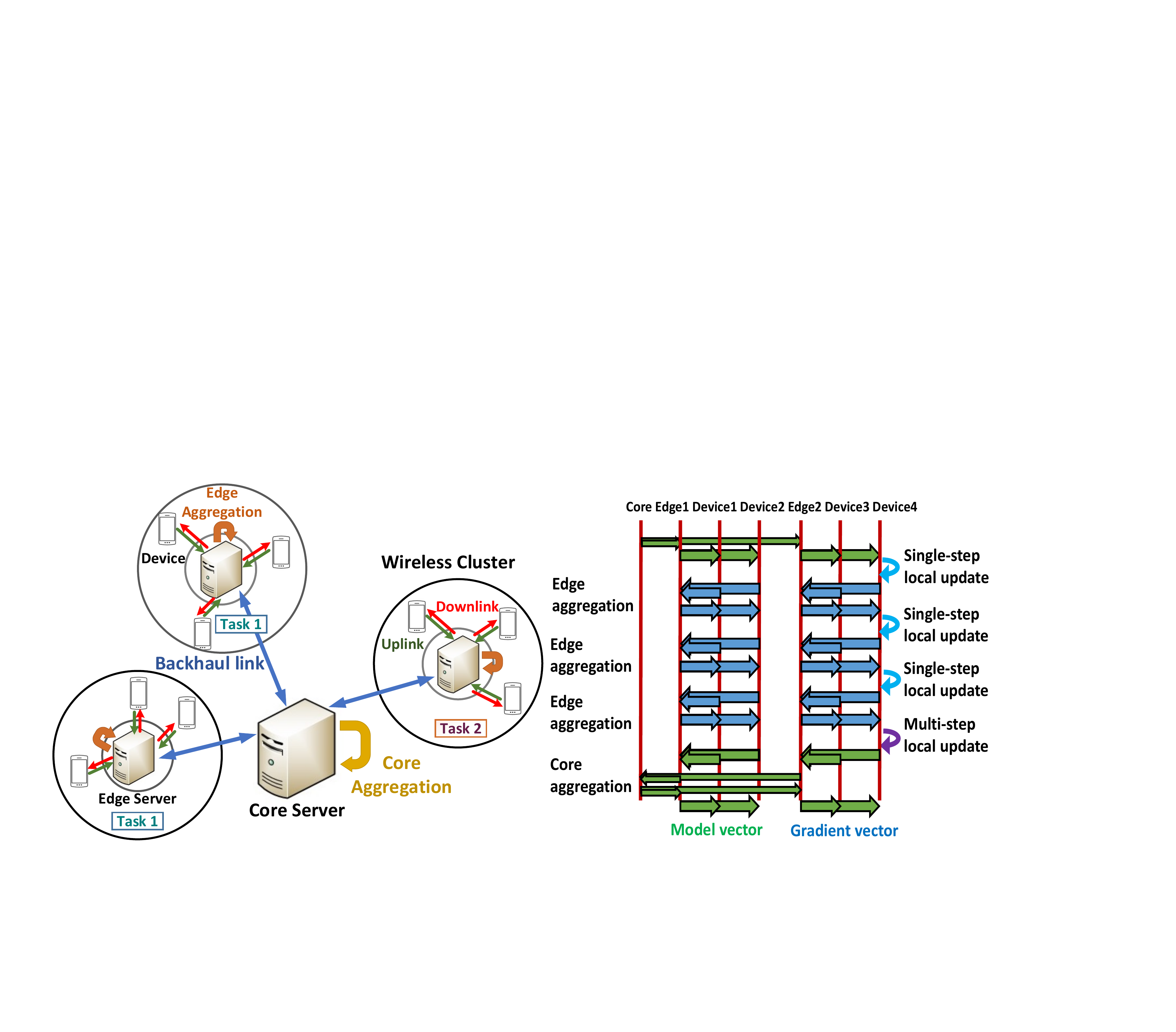}
	\vspace{-62pt}
	\caption{A hierarchical FL network, where three clusters with two learning tasks are illustrated, and {\fontfamily{lmtt}\selectfont
			MultiAirFed} method.}
	\vspace{-14pt}
\end{figure}

\textit{Network Topology:} We consider groups of devices clustered around edge servers, performing FL, as shown on Fig. 1. The clusters may have different or similar learning tasks. One or more core servers support the clusters in the learning process. For this, each edge server is connected to a core server by a backhaul link. The clusters with the same task connect to the same core server and collaborate to allow hierarchical learning.

We model the emerging network topology with the help of PCP \cite{haenggi_book}.
A PCP $\Phi$ is formally defined as a union of offspring points in $\mathbb{R}^2$ that are located around parent points. In our case, the parent points are the edge servers, while the offspring points are the devices.
The parent point process is a PPP $\Phi_\text{p}$ with density $\lambda_\text{p}$. The set of offspring points of $\mathbf{x} \in \Phi_\text{p}$ is denoted by cluster ${\cal N}^{{\mathbf{x}}}$, such that $\Phi = \cup_{\mathbf{x}\in \Phi_{\rm p}}{\cal N}^{{\mathbf{x}}}$. The PDF of each element of ${\cal N}^{{\mathbf{x}}}$ being at a location $\mathbf{y}+\mathbf{x} \in \mathbb{R}^2$ is shown by $f_{\|\mathbf{y}\|}(y)$. 
When disk-shaped clusters are employed in a PCP, the resulting point process is
referred to as Mat{\'e}rn cluster process (MCP) \cite{haenggi_book}. 
In addition, to model the protective zones around the antenna towers, that help to suppress interference by inhibiting nearby devices, we
further consider a modified type of MCP, named MCP with holes at the cluster centers (MCP-H) \cite{hole}, where the points are distributed
around cluster centers with uniform distribution inside rings with inner radius $r_0$ and outer radius $R$ as
$f_{\|\mathbf{y}\|}(y) = \frac{2y}{R^2-r_0^2}, \ r_0\leq y \leq R.$

The number of devices in each cluster is assumed to be $M$, i.e., $|{\cal N}^{{\mathbf{x}}}| = M$. Also, among all the devices of a cluster $\mathbf{x}$, the set of active devices in a time slot is denoted by $\cal A^{\mathbf{x}} \subseteq {\cal N}^{{\mathbf{x}}}$. The term "active device" denotes a device that participates in the aggregation phase of the FL by its uplink transmission. 

\textit{Channel Model:} All the nodes are single-antenna units. For the wireless links between devices and edge servers, we assume single-slope path loss and Rayleigh fading. The pathloss exponent is denoted by parameter $\alpha$. The uplink fading between a device $\mathbf{y} \in {\cal N}^{\mathbf{x}}$ and a server at $\mathbf{z}$ is modeled by $f_{\mathbf{y}\mathbf{z}}^\mathbf{x} \in \mathbb{C}$, such that $|f_{\mathbf{y}\mathbf{z}}^\mathbf{x}|^2 \sim \exp(1)$. The downlink fading between a server $\mathbf{x}$ and a reference device is $f^\mathbf{x}$ with the gain $|f^\mathbf{x}|^2 \sim \exp(\sigma_\text{d}^2)$, where $\sigma_\text{d}^2 >1$ is to model stronger downlink channels. 
The communication between the edge servers and the respective core server is considered error free.
\vspace{-4pt}
\section{Proposed Learning Method}
Assume that there are $C$ collaborating clusters including a reference cluster with its center at the origin $\mathbf{o}$ that have a same learning task. The centers of these clusters are denoted by a set $\mathcal{C}$. We propose a new hierarchical algorithm named {\fontfamily{lmtt}\selectfont
	MultiAirFed}, as a combination of intra-cluster gradient and inter-cluster
model parameter aggregation. Gradient aggregation has been shown to be robust to noise and interference in
\cite{huang_sg, yang3}, and to non-i.i.d. data in \cite{fedsgd-is-better}, and therefore is a good candidate for the intra-cluster learning process over the interfering wireless links.
The model-parameter aggregation at the core server at the same time allows multiple inter-cluster iterations. The algorithm is as follows: Consider a learning model with parameter vector $\mathbf{w} \in \mathbb{R}^d$, where $d$ denotes the learning model size. Let $T$ be the number of global inter-cluster iterations. In a iteration $t$, consider $\tau$ intra-cluster iterations. In a intra-cluster iteration $i$, each device $\mathbf{y}$ in a cluster $\mathbf{x}$ computes the local gradient of its loss function $F_{\mathbf{y}}^{\mathbf{x}}$ from its local dataset, indexed by $\left\{i,t\right\}$, as
\begin{align}
\label{localgrad}
\mathbf{g}_{\mathbf{y},i,t}^{\mathbf{x}} = \nabla F_{\mathbf{y}}^{\mathbf{x}}(\mathbf{w}_\mathbf{y}^\mathbf{x},\xi_\mathbf{y}^\mathbf{x}),
\end{align}
where $\mathbf{w}_\mathbf{y}^\mathbf{x}$ is its parameter vector, and ${\xi}_{\mathbf{y}}^{\mathbf{x}}$ with a cardinality of $B$ is the mini-batch
randomly chosen from its local dataset ${\cal D}_{\mathbf{y}}^{\mathbf{x}}$. Then, devices upload (transmit) their local gradients to their servers for intra-cluster aggregation. The server of cluster $\mathbf{x}$ averages of the local gradients from its active devices and broadcasts the generated intra-cluster gradient
\begin{align}
\label{intragrad}
\mathbf{g}_{i,t}^{{\mathbf{x}}} = \frac{1}{|{\cal A}_{i,t}^\mathbf{x}|}\sum_{\mathbf{y} \in {\cal A}_{i,t}^\mathbf{x}}^{} \mathbf{g}_{\mathbf{y},i,t}^{\mathbf{x}},
\end{align}
where $|{\cal A}_{i,t}^\mathbf{x}|$ is the number of active devices in the cluster
$\mathbf{x}$ for the iteration index $\left\{i,t\right\}$. Then, the servers broadcast the intra-cluster gradients $\mathbf{g}_{i,t}^{\mathbf{x}}, \forall \mathbf{x}$ to their devices. Utilizing $\mathbf{g}_{i,t}^{\mathbf{x}}$, each device $\mathbf{y}$ in any cluster $\mathbf{x}$ updates its local model following
a one-step gradient descent as
\begin{align}
\label{localcomp}
\mathbf{w}_{\mathbf{y},i+1,t}^{\mathbf{x}} = \mathbf{w}_{\mathbf{y},i,t}^{\mathbf{x}} -\mu_{t} \mathbf{g}_{i,t}^{\mathbf{x}},
\end{align}
where $\mu_{t}$ is the learning rate at the global iteration $t$. After completing $\tau$ intra-cluster iterations, each device performs a $\gamma$-step gradient descent locally as $
\mathbf{w}_{\mathbf{y},\tau,0,t}^{\mathbf{x}} = \mathbf{w}_{\mathbf{y},\tau,t}^{\mathbf{x}}$ and for $j = \left\{1,\cdots,\gamma\right\}$
\vspace{-5pt}
\begin{align}
\label{final_local01}
\mathbf{w}_{\mathbf{y},\tau,j,t}^{\mathbf{x}} = \mathbf{w}_{\mathbf{y},\tau,j-1,t}^{\mathbf{x}} - \mu_{t}\nabla F_{\mathbf{y}}^{\mathbf{x}}(\mathbf{w}_{\mathbf{y},\tau,j-1,t}^{\mathbf{x}},\xi_{\mathbf{y},\tau,j-1,t}^\mathbf{x}).
\end{align} 
To start the inter-cluster iteration, the devices upload their model parameters, i.e., $\mathbf{w}_{\mathbf{y},\tau,\gamma,t}^\mathbf{x}, \forall \mathbf{y}, \mathbf{x}$, to their servers. Accordingly, each server $\mathbf{x}$ computes an intra-cluster model parameter vector with the following average
\begin{align}
\label{intergrad}
\mathbf{w}_{t+1}^{{\mathbf{x}}} = \frac{1}{|{\cal A}_{\tau,t}^\mathbf{x}|}\sum_{\mathbf{y} \in {\cal A}_{\tau,t}^\mathbf{x}}^{} \mathbf{w}_{\mathbf{y},\tau,\gamma,t}^{\mathbf{x}}.
\end{align}
Then, collaborating servers upload their intra-cluster model parameter vectors to the core server for a global inter-cluster model parameter aggregation as
\begin{align}
\label{globalgrad}
\mathbf{w}_{t+1}^{\text{G}} = \frac{1}{|{\cal A}_{\tau,t}|} \sum_{\mathbf{x} \in {\mathcal{C}}}^{} |{\cal A}_{\tau,t}^{\mathbf{x}}| \mathbf{w}_{t+1}^{\mathbf{x}},
\end{align}
where $|{\cal A}_{\tau,t}| = \sum_{\mathbf{x}\in{\mathcal{C}}}^{}|{\cal A}_{\tau,t}^\mathbf{x}|$. Then, the servers broadcast $\mathbf{w}_{t+1}^{\text{G}}$ to the devices to update their initial state of model parameter vector for the next global iteration $t+1$ as $\mathbf{w}_{\mathbf{y},0,t+1}^{\mathbf{x}} =  \mathbf{w}_{t+1}^{\text{G}}, \forall \mathbf{y}, \mathbf{x}$. This global update synchronizes all the devices in the collaborating clusters and prevents a high deviation of the local training processes. An illustration of this algorithm is in Fig. 1. 

Compared to other hierarchical methods in \cite{letaief, bennis, tony, gunduz_ota} which are based on model parameter transmissions, gradient transmission in {\fontfamily{lmtt}\selectfont
	MultiAirFed} over wireless links is expected to be more robust to channel noise. This is because for each device local update, a noisy model parameter aggregation leads to imperfections on both the initial state update and the local gradient function evaluation in \eqref{localgrad}. Moreover, the resulting errors propagate and reinforce through multiple local steps. As learning convergence demands high accuracy in the gradient direction, particularly in the vicinity of the optimal solution, noisy model parameter aggregation may hinder the model to converge. However, when gradients are transmitted, devices can download the aggregated gradient \eqref{intragrad} as a same gradient term for their update without the need for local computations, and the initial state remains unaffected by any noise. Furthermore, in general, gradient based aggregation is shown to be resilient against heterogeneity and non-i.i.d. data when compared to approaches that transmit model parameters \cite{fedsgd-is-better}. Although gradient transmission allows only one local iteration per intra-cluster iteration, the proposed gradient descent \eqref{final_local01} reinforces the inclusion of local training in the learning process. These will be further justified via experimental results in Section V.

Please note that the proposed learning method is not limited to the choice of the spatial model in Subsection II.A and the sequel transmission scheme in Section IV.

\vspace{-4pt}
\section{Transmission Scheme}
To implement {\fontfamily{lmtt}\selectfont
	MultiAirFed}, we propose a scalable transmission scheme including two types of analog transmissions for uplink and downlink, where each is done simultaneously over the clusters in a single resource block. It is inspired from \cite{downlink} which shows that analog downlink approach significantly outperforms the digital one. 
From here, we ignore the iteration indexes for simplicity of presentation.

\textit{Uplink:} Depending on an intra- or inter-cluster iteration, the gradient parameters or model parameters at each
device are normalized before transmission to have zero mean and unit variance. Normalizing the parameters offers two benefits. First, when the parameters have zero-mean entries, the estimates obtained in the sequel are unbiased. Second, when the entries have unit variance, the interference and consequently the error terms do not depend on the specific values of parameters. 

Due to data heterogeneity, devices can exhibit varying mean and variance. For an intra-cluster iteration, the local gradient vector at a device $\mathbf{y} \in {\cal N}^{\mathbf{x}}$, i.e., ${{\mathbf{g}}_{\mathbf{y}}^\mathbf{x}}$, is normalized as ${\bar{\mathbf{g}}_{\mathbf{y}}^\mathbf{x}} = \frac{{{\mathbf{g}}_{\mathbf{y}}^\mathbf{x}}-\mu_{\text{g},\mathbf{y}}^\mathbf{x}\mathbf{1}}{\sigma_{\text{g},\mathbf{y}}^\mathbf{x}}$, where $\mathbf{1}$ is the all one vector, and $\mu_{\text{g},\mathbf{y}}^\mathbf{x}$ and ${\sigma_{\text{g},\mathbf{y}}^\mathbf{x}}$ denote the mean and standard deviation of the $d$ entries of the gradient given by $
\mu_{\text{g},\mathbf{y}}^\mathbf{x} = \frac{1}{d}\sum_{i=1}^{d} {\mathbf{g}}_\mathbf{y}^\mathbf{x}(i),\
\sigma_{\text{g},\mathbf{y}}^{\mathbf{x}\ 2} = \frac{1}{d}\sum_{i=1}^{d}({\mathbf{g}}_\mathbf{y}^\mathbf{x}(i)-\mu_{\text{g},\mathbf{y}}^\mathbf{x})^2,$
where $\mathbf{g}_\mathbf{y}^\mathbf{x}(i)$ is the $i$-th entry of the vector.
Also, for an inter-cluster iteration, the normalized local model parameter vector is ${\bar{\mathbf{w}}_{\mathbf{y}}^\mathbf{x}} = \frac{{{\mathbf{w}}_{\mathbf{y}}^\mathbf{x}}-\mu_{\text{w},\mathbf{y}}^\mathbf{x}\mathbf{1}}{\sigma_{\text{w},\mathbf{y}}^\mathbf{x}}$, where the mean and variance are $
\mu_{\text{w},\mathbf{y}}^\mathbf{x} = \frac{1}{d}\sum_{i=1}^{d} {\mathbf{w}}_\mathbf{y}^\mathbf{x}(i),\
\sigma_{\text{w},\mathbf{y}}^{\mathbf{x}\ 2} = \frac{1}{d}\sum_{i=1}^{d}({\mathbf{w}}_\mathbf{y}^\mathbf{x}(i)-\mu_{\text{w},\mathbf{y}}^\mathbf{x})^2.$
Then, at each device $\mathbf{y}$ in the cluster $\mathbf{x}$, the normalized vector ${\bar{\mathbf{g}}_{\mathbf{y}}^\mathbf{x}}$ or ${\bar{\mathbf{w}}_{\mathbf{y}}^\mathbf{x}}$ is analog modulated and
transmitted as ${p_{\mathbf{y}}^{\mathbf{x}}} {\bar{\mathbf{g}}_{\mathbf{y}}^\mathbf{x}}$ or ${p_{\mathbf{y}}^{\mathbf{x}}} {\bar{\mathbf{w}}_{\mathbf{y}}^\mathbf{x}}$ simultaneously with other devices in all the clusters, where $|{p_{\mathbf{y}}^{\mathbf{x}}}|^2$ denotes the transmission power. Thus, the received signal at a server located at $\mathbf{z}$ is $\mathbf{v}_\text{u}^{\mathbf{z}} =$ 
\begin{align}
\label{uplink}
&\sum_{\mathbf{y} \in {\cal N}^{\mathbf{z}}}^{}{p_{\mathbf{y}}^{\mathbf{z}}} \|\mathbf{y}\|^{-\frac{\alpha}{2}} f_{\mathbf{y}\mathbf{z}}^\mathbf{z} {\bar{\mathbf{s}}_{\mathbf{y}}^\mathbf{z}}+\sum_{\mathbf{x} \in \Phi_\text{p}\backslash \left\{\mathbf{z}\right\}}^{}\sum_{\mathbf{y} \in {\cal N}^{\mathbf{x}}}^{} {p_{\mathbf{y}}^{\mathbf{x}}} \|\mathbf{x}+\mathbf{y}-\mathbf{z}\|^{-\frac{\alpha}{2}} \nonumber\\
&\times f_{\mathbf{y}\mathbf{z}}^\mathbf{x} {\bar{\mathbf{s}}_{\mathbf{y}}^\mathbf{x}}, \ \bar{\mathbf{s}}_{\mathbf{y}}^\mathbf{x} = \left\{{\bar{\mathbf{g}}_{\mathbf{y}}^\mathbf{x}},{\bar{\mathbf{w}}_{\mathbf{y}}^\mathbf{x}}\right\},
\end{align}
where the first term is the useful signal and the second is the inter-cluster interference. In \eqref{uplink}, we ignore the receiver noise compared to the interference. Each device $\mathbf{y} \in {\cal N}^\mathbf{x}$ of cluster $\mathbf{x}$ follows a truncated power allocation scheme \cite{huang_analog, huang_sg} as $p_\mathbf{y}^{\mathbf{x}} = 
\frac{\sqrt{\rho}}{\|\mathbf{y}\|^{-\frac{\alpha}{2}}f_{\mathbf{y}\mathbf{x}}^{\mathbf{x}}}$ if $|f_{\mathbf{y}\mathbf{x}}^{\mathbf{x}}|^2 \geq \text{th}_1$, and $p_\mathbf{y}^{\mathbf{x}} = 0$ if $|f_{\mathbf{y}\mathbf{x}}^{\mathbf{x}}|^2 < \text{th}_1$, where $\rho$ is the power allocation parameter and $\text{th}_1$ is a threshold. We assume that the device knows this channel, the uplink channel to its server. In the power allocation, to meet a maximum average power $P_\text{u}$ in each device, we have $\mathbb{E}\left\{|p_\mathbf{y}^\mathbf{x}|^2\right\} = \mathbb{E}\left\{\frac{\rho}{\|\mathbf{y}\|^{-\alpha}|f_{\mathbf{y}\mathbf{x}}^{\mathbf{x}}|^2}\right\} = \rho \mathbb{E}\biggl\{\left.\frac{1}{|f_{\mathbf{y}\mathbf{x}}^{\mathbf{x}}|^2}\right \vert |f_{\mathbf{y}\mathbf{x}}^{\mathbf{x}}|^2 > \text{th}_1 \biggr\} \mathbb{E}\left\{\|\mathbf{y}\|^\alpha\right\} = \rho \text{Ei}(\text{th}_1) \int_{r_0}^{R} \frac{2y^{1+\alpha}}{R^2-r_0^2} \mathrm{d}y = \frac{2\rho}{2+\alpha} \text{Ei}(\text{th}_1) \frac{R^{\alpha+2}-r_0^{\alpha+2}}{R^2-r_0^2} \leq P_\text{u},$
where $\text{Ei}(x) =\int_{x}^{\infty}\frac{e^{-t}}{t}\mathrm{d}t$ is the exponential integral function. Thus, $\rho$ for all the devices can be selected as
$\rho = \frac{(2+\alpha)(R^2-r_0^2)}{2 \text{Ei}(\text{th}_1)(R^{\alpha+2}-r_0^{\alpha+2})}P_\text{u}.$ 

\textit{Downlink for intra-cluster iteration:} As $\mathbb{E}\left\{\mathbf{v}_\text{u}^{\mathbf{x}}\right\} = \mathbf{0}, \forall \mathbf{x}$, each server at a location $\mathbf{x} \in \Phi_\text{p}$ normalizes its received signal $\mathbf{v}_\text{u}^{\mathbf{x}}$ with its variance, which is $\mathbb{E}\left\{\|\mathbf{v}_\text{u}^{\mathbf{x}}\|^2\right\}$, as $\frac{\mathbf{v}_\text{u}^{\mathbf{x}}}{\sqrt{\mathbb{E}\left\{\|\mathbf{v}_\text{u}^{\mathbf{x}}\|^2\right\}}}$. Then, all the servers transmit the normalized
signals simultaneously. Therefore, the received signal at a reference device at $\mathbf{y}_0$ in the reference cluster $\mathbf{o}$ is
\begin{align}
\label{intradownlink}
\mathbf{v}_{\text{d}_{\mathbf{y}_0}}^{\mathbf{o}} = \sum_{\mathbf{x} \in \Phi_\text{p}}^{} \sqrt{\frac{{P_{\text{d}}}}{{\mathbb{E}\left\{\|\mathbf{v}_\text{u}^{\mathbf{x}}\|^2\right\}}}} \|\mathbf{x}+\mathbf{y}_0\|^{-\frac{\alpha}{2}} f^\mathbf{x} {\mathbf{v}_\text{u}^{\mathbf{x}}},
\end{align}
where $P_{\text{d}}$ is the transmission power constraint of the servers. In general, the server can estimate $\mathbb{E}\left\{\|\mathbf{v}_\text{u}^\mathbf{x}\|^2\right\}$ by taking measurements of the received signal over time and its entries and calculating the average power of those samples. However, the MCP-H modeling allows us to express $\mathbb{E}\left\{\|\mathbf{v}_\text{u}^\mathbf{x}\|^2\right\}$  as a function of the network parameters and the power control. Specifically, from \eqref{uplink} and the power allocation
\begin{align}
\label{power}
\mathbb{E}\left\{\|\mathbf{v}_\text{u}^\mathbf{x}\|^2\right\} = \rho|{\cal A}^\mathbf{x}|+\Psi,
\end{align}
where $\rho |{\cal A}^\mathbf{x}|$ is the received power from the active devices in the cluster $\mathbf{x}$ and $\Psi$ is the inter-cluster uplink interference power from other clusters given in the next theorem.
\begin{theorem}
	The inter-cluster uplink interference power is
	\begin{align}
	\label{upper_error_derive}
	&\Psi = 2M \rho \lambda_\text{p} \frac{e^{-\text{th}_1} \text{Ei}(\text{th}_1)}{R^2-r_0^2}\int_{2r_0}^{\infty}\int_{0}^{2\pi}\int_{r_0}^{R}  y\biggl(1+\frac{x^2}{y^2}-2\frac{x}{y}\nonumber\\
	&\times\cos(\theta)\biggr)^{-\frac{\alpha}{2}} x \mathrm{d}y \mathrm{d}\theta \mathrm{d}x.\hspace{-100pt}
	\end{align}
\end{theorem}
\begin{IEEEproof}
	See Appendix A.
\end{IEEEproof}

The result in Theorem 1 will be further utilized subsequently for the analysis of uplink and downlink aggregation errors.

Denormalizing received signal, the reference device estimates the aggregated intra-cluster gradient \eqref{intragrad} as ${\mathbf{g}}_{\mathbf{y}_0}^\mathbf{o} =$
\begin{align}
\label{intraestimate1}
\frac{ \vartheta_{\text{d}_{\mathbf{y}_0}}^{\mathbf{o}} \mathbf{v}_{\text{d}_{\mathbf{y}_0}}^{\mathbf{o}}}{\sqrt{\rho} \sqrt{\frac{{P_{\text{d}}}}{{\mathbb{E}\left\{\|\mathbf{v}_\text{u}^{\mathbf{o}}\|^2\right\}}}}|{\cal A}^\mathbf{o}|f^{\mathbf{o}}\|\mathbf{y}_0\|^{-\frac{\alpha}{2}}} +\frac{1}{|{\cal A}^\mathbf{o}|}\sum_{\mathbf{y} \in {\cal A}^\mathbf{o}} \mu_{\text{g},\mathbf{y}}^\mathbf{o}\mathbf{1},
\end{align}
where $\vartheta_{\text{d}_{\mathbf{y}_0}}^{\mathbf{o}}$ is the intra-cluster receive normalizing factor at the device. For this operation, it is assumed that each device knows its downlink channel from its server and the reference server shares the scalars $({\mu_{\text{g},\mathbf{y}}^\mathbf{o}}, {\sigma_{\text{g},\mathbf{y}}^\mathbf{o}}), \forall \mathbf{y} \in {\cal A}^\mathbf{o}$ with its devices in an error-free manner. This is needed to support data heterogeneity. By replacing \eqref{uplink} in \eqref{intradownlink}, \eqref{intraestimate1} can be rewritten as 
\begin{align}
\label{intraestimate}
&{\mathbf{g}}_{\mathbf{y}_0}^\mathbf{o} = \frac{1}{|{\cal A}^\mathbf{o}|} \sum_{\mathbf{y}\in {\cal A}^\mathbf{o}}^{} {\mathbf{g}}_{\mathbf{y}}^\mathbf{o}+\frac{ \vartheta_{\text{d}_{\mathbf{y}_0}}^{\mathbf{o}} \mathbf{v}_{\text{d}_{\mathbf{y}_0}}^{\mathbf{o}}}{\sqrt{\rho} \sqrt{\frac{{P_{\text{d}}}}{{\mathbb{E}\left\{\|\mathbf{v}_\text{u}^{\mathbf{o}}\|^2\right\}}}}|{\cal A}^\mathbf{o}|f^{\mathbf{o}}\|\mathbf{y}_0\|^{-\frac{\alpha}{2}}}-\nonumber\\& \frac{\sum_{\mathbf{y} \in {\cal A}^\mathbf{o}} \left({\mathbf{g}}_{\mathbf{y}}^\mathbf{o}-\mu_{\text{g},\mathbf{y}}^\mathbf{o}\mathbf{1}\right)}{|{\cal A}^\mathbf{o}|} =\frac{\sum_{\mathbf{y}\in {\cal A}^\mathbf{o}}^{} {\mathbf{g}}_{\mathbf{y}}^\mathbf{o}}{|{\cal A}^\mathbf{o}|} + \frac{ \boldsymbol{\epsilon}_\text{u}^\mathbf{o}}{ |{\cal A}^{\mathbf{o}}|} +\frac{\boldsymbol{\epsilon}_{\text{d}_{\mathbf{y}_0}}^{\mathbf{o}}}{|{\cal A}^\mathbf{o}|},
\end{align}
where $\boldsymbol{\epsilon}_\text{u}^\mathbf{o}$ is the intra-cluster uplink error, from the interference of devices
in the uplink, given by 
\begin{align}
\label{intrauplinkerror}
&\sum_{\mathbf{y} \in {\cal A}^\mathbf{o}}\left(\vartheta_{\text{d}_{\mathbf{y}_0}}^{\mathbf{o}}-\sigma_{\text{g},\mathbf{y}}^\mathbf{o}\right)\bar{\mathbf{g}}_{\mathbf{y}}^\mathbf{o}+\nonumber\\&\vartheta_{\text{d}_{\mathbf{y}_0}}^{\mathbf{o}}\sum_{\mathbf{x}\in \Phi_\text{p}}\sum_{\mathbf{y} \in {\cal N}^{\mathbf{x}}} \mathds{1}\left(|f_{\mathbf{y}\mathbf{x}}^{\mathbf{x}}|^2>\text{th}_1\right) \frac{\|\mathbf{x}+\mathbf{y}\|^{-\frac{\alpha}{2}}}{\|\mathbf{y}\|^{-\frac{\alpha}{2}}} \frac{f_{\mathbf{y}\mathbf{o}}^\mathbf{x}}{f_{\mathbf{y}\mathbf{x}}^{\mathbf{x}}}\bar{\mathbf{g}}_{\mathbf{y}}^\mathbf{x},
\end{align}
and the intra-cluster downlink error $\boldsymbol{\epsilon}_{\text{d}_{\mathbf{y}_0}}^{\mathbf{o}}$, from the interference of servers in the downlink, is
\begin{align}
\label{intradownlinkerror}
&\frac{\vartheta_{\text{d}_{\mathbf{y}_0}}^{\mathbf{o}}\sum_{\mathbf{x} \in \Phi_\text{p}}^{} \sqrt{\frac{{P_{\text{d}}}}{{\mathbb{E}\left\{\|\mathbf{v}_\text{u}^{\mathbf{x}}\|^2\right\}}}} \|\mathbf{x}+\mathbf{y}_0\|^{-\frac{\alpha}{2}} f^\mathbf{x} {\mathbf{v}_\text{u}^{\mathbf{x}}}}{\sqrt{\rho} \sqrt{\frac{{P_{\text{d}}}}{{\mathbb{E}\left\{\|\mathbf{v}_\text{u}^{\mathbf{o}}\|^2\right\}}}}f^{\mathbf{o}}\|\mathbf{y}_0\|^{-\frac{\alpha}{2}}}.
\end{align}
These results hold for the learning process under general network topology. For the specific case of the MCP-H, we can select the normalizing factor $\vartheta_{\text{d}_{\mathbf{y}_0}}^{\mathbf{o}}$ in \eqref{intraestimate1} to minimize the distortion of the recovered gradient ${\mathbf{g}}_{\mathbf{y}_0}^\mathbf{o}$ with respect to the ground true gradient $\frac{1}{|{\cal A}^\mathbf{o}|} \sum_{\mathbf{y}\in {\cal A}^\mathbf{o}}^{} {\mathbf{g}}_{\mathbf{y}}^\mathbf{o}$ from \eqref{intraestimate}, which can be measured by the mean squared error (MSE)\cite{viktoria, huang_sg}, as
\begin{align}
\label{factor_opt}
&\min_{\vartheta_{\text{d}_{\mathbf{y}_0}}^{\mathbf{o}}} \mathbb{E}\left\{\left\Vert\frac{ \boldsymbol{\epsilon}_\text{u}^\mathbf{o}}{|{\cal A}^\mathbf{o}|} +\frac{\boldsymbol{\epsilon}_{\text{d}_{\mathbf{y}_0}}^{\mathbf{o}}}{|{\cal A}^\mathbf{o}|}\right\Vert^2\right\} \nonumber\\
&=\frac{1}{|{\cal A}^\mathbf{o}|^2}\left(\mathbb{E}\left\{\|\boldsymbol{\epsilon}_\text{u}^\mathbf{o}\|^2\right\}+\mathbb{E}\left\{\|\boldsymbol{\epsilon}_{\text{d}_{\mathbf{y}_0}}^\mathbf{o}\|^2\right\}\right),
\end{align}
where the equality holds due to the independent error terms and $\mathbb{E}\left\{\|\boldsymbol{\epsilon}_\text{u}^\mathbf{o}\|^2\right\} = \sum_{\mathbf{y} \in {\cal A}^\mathbf{o}}\left(\vartheta_{\text{d}_{\mathbf{y}_0}}^{\mathbf{o}}-\sigma_{\text{g},\mathbf{y}}^\mathbf{o}\right)^2+\frac{{\vartheta_{\text{d}_{\mathbf{y}_0}}^{\mathbf{o}\ \hspace{-2pt}2}}}{\rho} \Psi$ from \eqref{uplink} and \eqref{intrauplinkerror}, where $\Psi$ is calculated in \eqref{upper_error_derive}. Then, for the expected term on the intra-cluster downlink error in \eqref{intradownlinkerror}, due to the MCP-H network topology, we have
\begin{align}
\label{downlink_error_derive}
&\mathbb{E}\left\{\|\boldsymbol{\epsilon}_{\text{d}_{\mathbf{y}_0}}^\mathbf{o}\|^2\right\}  = \frac{{\vartheta_{\text{d}_{\mathbf{y}_0}}^{\mathbf{o}\ \hspace{-2pt}2}}\mathbb{E}\left\{\sum_{\mathbf{x} \in \Phi_\text{p}}^{} {{{P_{\text{d}}}}} \|\mathbf{x}+\mathbf{y}_0\|^{-{\alpha}} |f^\mathbf{x}|^2\right\}}{{\rho} {\frac{{P_{\text{d}}}}{{\mathbb{E}\left\{\|\mathbf{v}_\text{u}^{\mathbf{o}}\|^2\right\}}}}|f^{\mathbf{o}}|^2\|\mathbf{y}_0\|^{-{\alpha}}}   \stackrel{(a)}{=}\nonumber\\
&\frac{{\vartheta_{\text{d}_{\mathbf{y}_0}}^{\mathbf{o}\ \hspace{-2pt}2}}\mathbb{E}\left\{\|\mathbf{v}_\text{u}^\mathbf{o}\|^2\right\}}{\rho|f^\mathbf{o}|^2\|\mathbf{y}_0\|^{-\alpha}}  \left(\sigma_\text{d}^2\times 2\pi \lambda_\text{p} \int_{r_0}^{\infty} x^{1-\alpha}\mathrm{d}x \right)\stackrel{(b)}{=} \nonumber\\
& \frac{{\vartheta_{\text{d}_{\mathbf{y}_0}}^{\mathbf{o}\ \hspace{-2pt}2}}}{\rho|f^\mathbf{o}|^2\|\mathbf{y}_0\|^{-\alpha}} \left(\frac{2\pi \lambda_\text{p}\sigma_\text{d}^2}{(\alpha-2)r_0^{\alpha-2}}\right) (\rho|{\cal A}^\mathbf{o}|+\Psi),
\end{align}
where $(a)$ is due to the Campbell's theorem \cite{haenggi_book} and $(b)$ is from \eqref{power}. The distortion in \eqref{factor_opt} is an important component in the convergence analysis of {\fontfamily{lmtt}\selectfont MultiAirFed}, as given in \cite{journal}, such that a reduction in distortion leads to a significant decrease in the optimality gap.

The solution of \eqref{factor_opt} is given in the next theorem. 
\begin{theorem}
	The optimal normalizing factor $\vartheta_{\text{d}_{\mathbf{y}_0}}^{\mathbf{o}}$ is
	\begin{align}
	\label{intra_dowlik_norm}
	\vartheta_{\text{d}_{\mathbf{y}_0}}^{\mathbf{o}} =\frac{\sum_{\mathbf{y} \in {\cal A}^\mathbf{o}}^{} \sigma_{\text{g},\mathbf{y}}^\mathbf{o}}{\left(1+\frac{\beta}{|f^\mathbf{o}|^2\|\mathbf{y}_0\|^{-\alpha}}\right)\left(|{\cal A}^\mathbf{o}|+\frac{\Psi}{\rho}\right)},
	\end{align}
	where $\beta = \frac{2\pi \lambda_\text{p}\sigma_\text{d}^2}{(\alpha-2)r_0^{\alpha-2}}$.
\end{theorem}
\begin{IEEEproof}
	To solve \eqref{factor_opt}, we take derivative from the objective and set the result to zero, which leads to $\sum_{\mathbf{y} \in {\cal A}^\mathbf{o}}\left(\vartheta_{\text{d}_{\mathbf{y}_0}}^{\mathbf{o}}-\sigma_{\text{g},\mathbf{y}}^\mathbf{o}\right)+\frac{{\vartheta_{\text{d}_{\mathbf{y}_0}}^{\mathbf{o}}}}{\rho} \Psi+\frac{{\vartheta_{\text{d}_{\mathbf{y}_0}}^{\mathbf{o}}}}{|f^\mathbf{o}|^2\|\mathbf{y}_0\|^{-\alpha}} \beta (|{\cal A}^\mathbf{o}|+\frac{\Psi}{\rho}) = 0$, resulting in \eqref{intra_dowlik_norm}.
\end{IEEEproof}	
Replacing \eqref{intra_dowlik_norm}, the minimum distortion in \eqref{factor_opt} is obtaind as
\begin{align}
\label{remark}
\hspace{-10pt}\frac{\sum_{\mathbf{y} \in {\cal A}^\mathbf{o}}^{} \sigma_{\text{g},\mathbf{y}}^{\mathbf{o}\ \hspace{2pt}2}}{|{\cal A}^\mathbf{o}|^2} - \frac{(\sum_{\mathbf{y} \in {\cal A}^\mathbf{o}}^{} \sigma_{\text{g},\mathbf{y}}^{\mathbf{o}})^2}{\left(1+\frac{\beta}{|f^\mathbf{o}|^2\|\mathbf{y}_0\|^{-\alpha}}\right)|{\cal A}^\mathbf{o}|^2\left(|{\cal A}^\mathbf{o}|+\frac{\Psi}{\rho}\right)}
\end{align}
\begin{remark}
	Since $\beta$ and $\Psi$ are linearly dependent to $\lambda_\text{p}$, the distortion of the proposed intra-cluster aggregation in \eqref{intraestimate1} and its effect on the learning performance increases with the order $\mathcal{O}\left(\text{const}-\frac{1}{\lambda_\text{p}^2}\right)$. In the absence of the optimal normalizing factor, the order is $\mathcal{O}\left({\lambda_\text{p}^2}\right)$ from \eqref{factor_opt} and \eqref{downlink_error_derive}. This shows that the normalization factor can significantly diminish the impact of uplink and downlink interferences, where the number of interfering devices and servers are scaled by cluster density $\lambda_\text{p}$, respectively.
\end{remark}

In the case of i.i.d. data distribution over devices, i.e., $\sigma_{\text{g},\mathbf{y}}^{\mathbf{o}}=\sigma_{\text{g},\mathbf{y}_0}^{\mathbf{o}}, \forall \mathbf{y} \in {\cal A}^\mathbf{o}$, the minimum distortion \eqref{remark} is simplified to 
\begin{align}
\frac{\sigma_{\text{g},\mathbf{y}_0}^{\mathbf{o}\ \hspace{2pt}2}}{|{\cal A}^\mathbf{o}|+\rho \frac{|{\cal A}^\mathbf{o}|^2}{\Psi}}+\frac{\sigma_{\text{g},\mathbf{y}_0}^{\mathbf{o}\ \hspace{2pt}2} \beta}{(|f^\mathbf{o}|^2\|\mathbf{y}_0\|^{-\alpha}+\beta)\left(|{\cal A}^\mathbf{o}|+\frac{\Psi}{\rho}\right)},
\end{align}
where the first part is the effect of uplink interference and the second part is from the downlink interference.
\begin{remark}
	Considering the i.i.d. scenario, increasing the number of active devices within the reference cluster has the effect of reducing the distortion resulting from uplink interference, with an order $\mathcal{O}\left(\frac{1}{|{\cal A}^\mathbf{o}|^2}\right)$. Similarly, the distortion resulting from downlink interference is also reduced, but with an order $\mathcal{O}\left(\frac{1}{|{\cal A}^\mathbf{o}|}\right)$. Therefore, the scaling impact of $|{\cal A}^\mathbf{o}|$ on performance improvement is attenuated by one exponent in the presence of downlink interference.
\end{remark} 

\textit{Downlink for inter-cluster iteration:}
The core server sums and redistributes the signals received
from any set of collaborating edge servers. Consider $C_\mathbf{x}$ collaborating clusters having the same learning task with a cluster $\mathbf{x}$, denoted as the set ${\mathcal{C}}_{\mathbf{x}}$. Then, the sum of received signals of the clusters in ${\mathcal{C}}_{\mathbf{x}}$, i.e., $\sum_{\mathbf{z} \in {\mathcal{C}}_\mathbf{x}}^{}\mathbf{v}_\text{u}^{\mathbf{z}}$, is normalized with its variance, which is $\sum_{\mathbf{z} \in {\mathcal{C}}_\mathbf{x}}^{}\mathbb{E}\left\{\|\mathbf{v}_\text{u}^{\mathbf{z}}\|^2\right\}$, as $\frac{\sum_{\mathbf{z} \in {\mathcal{C}}_\mathbf{x}}^{}\mathbf{v}_\text{u}^{\mathbf{z}}}{\sqrt{\sum_{\mathbf{z} \in {\mathcal{C}}_\mathbf{x}}^{}\mathbb{E}\left\{\|\mathbf{v}_\text{u}^{\mathbf{z}}\|^2\right\}}}$. Then, the result is simultaneously transmitted from the servers of the clusters in ${\mathcal{C}}_\mathbf{x}$ to their devices. Therefore, the received signal at the reference device is $\mathbf{v}_{\text{d}_{\mathbf{y}_0}} =$
\begin{align}
\label{interdownlink}
\sum_{\mathbf{x} \in \Phi_\text{p}}^{} \sqrt{\frac{{P_{\text{d}}}}{\sum_{\mathbf{z} \in {\mathcal{C}}_\mathbf{x}}^{}\mathbb{E}\left\{\|\mathbf{v}_\text{u}^{\mathbf{z}}\|^2\right\}}} \|\mathbf{x}+\mathbf{y}_0\|^{-\frac{\alpha}{2}} f^\mathbf{x} \sum_{\mathbf{z} \in {\mathcal{C}}_\mathbf{x}}^{}{\mathbf{v}_\text{u}^{\mathbf{z}}}.
\end{align}
Then, the reference device can estimate the aggregated inter-cluster model parameter vector as ${\mathbf{w}}_{\mathbf{y}_0}^\mathbf{o} =$
\vspace{-5pt}
\begin{align}
\label{interestimate1}
\hspace{-8pt}\frac{\vartheta_{\text{d}_{\mathbf{y}_0}} \mathbf{v}_{\text{d}_{\mathbf{y}_0}}}{\sqrt{\rho}\sqrt{\frac{{P_{\text{d}}}}{\sum_{\mathbf{x} \in {\mathcal{C}}}^{}\mathbb{E}\left\{\|\mathbf{v}_\text{u}^{\mathbf{x}}\|^2\right\}}} f^{\mathbf{o}}\|\mathbf{y}_0\|^{-\frac{\alpha}{2}}|{\cal A}|}\hspace{-2pt}+\hspace{-2pt}\frac{1}{|{\cal A}|}\sum_{\mathbf{x} \in {\cal C}}\hspace{-1pt}\sum_{\mathbf{y} \in {\cal A}^\mathbf{x}} \hspace{-3pt}\mu_{\text{w},\mathbf{y}}^\mathbf{x}\mathbf{1}\hspace{-2pt}
\end{align}
where due to the symmetry of the network $\sum_{\mathbf{x} \in {\mathcal{C}}}^{}\mathbb{E}\left\{\|\mathbf{v}_\text{u}^{\mathbf{x}}\|^2\right\} = \rho|{\cal A}|+C{\Psi}$ and $|{\cal A}| = \sum_{\mathbf{x}\in {\mathcal{C}}}^{}|{\cal A}^\mathbf{x}|$. Also, $\vartheta_{\text{d}_{\mathbf{y}_0}}$ is the inter-cluster receive normalizing factor. We assume that the reference server shares scalars $( {\mu_{\text{w},\mathbf{y}}^\mathbf{x}}, {\sigma_{\text{w},\mathbf{y}}^\mathbf{x}}), \forall \mathbf{x} \in {\cal C}, \mathbf{y}\in {{\cal A}^\mathbf{x}} $ among its devices. After replacing \eqref{uplink} in \eqref{interdownlink}, \eqref{interestimate1} can be expanded as
\begin{align}
\label{interestimate}
&{\mathbf{w}}_{\mathbf{y}_0}^\mathbf{o}  = \frac{1}{|{\cal A}|} \sum_{\mathbf{x}\in {\mathcal{C}}}^{} \sum_{\mathbf{y}\in {\cal A}^{\mathbf{x}}}^{} {\mathbf{w}}_{\mathbf{y}}^\mathbf{x} + \frac{\boldsymbol{\epsilon}_\text{u}}{|\cal A|}+\frac{\boldsymbol{\epsilon}_{\text{d}_{\mathbf{y}_0}}}{|\cal A|},
\end{align}
where $\boldsymbol{\epsilon}_\text{u}$ is the inter-cluster uplink error as  
\begin{align}
\label{interuplinkerror}
&\sum_{\mathbf{x}\in {\cal C}}^{} \sum_{\mathbf{y} \in {\cal A}^\mathbf{x}}^{}\left(\vartheta_{\text{d}_{\mathbf{y}_0}}-\sigma_{\text{w},\mathbf{y}}^\mathbf{x}\right)\bar{\mathbf{w}}_\mathbf{y}^\mathbf{x}+ \vartheta_{\text{d}_{\mathbf{y}_0}} \sum_{\mathbf{z} \in {\mathcal{C}}}^{}\sum_{\mathbf{x}\in \Phi_\text{p}}\sum_{\mathbf{y} \in {\cal N}^{\mathbf{x}}} \nonumber\\
&\mathds{1}\left(|f_{\mathbf{y}\mathbf{x}}^{\mathbf{x}}|^2>\text{th}_1\right) \frac{\|\mathbf{x}+\mathbf{y}-\mathbf{z}\|^{-\frac{\alpha}{2}}}{\|\mathbf{y}\|^{-\frac{\alpha}{2}}} \frac{f_{\mathbf{y}\mathbf{z}}^\mathbf{x}}{f_{\mathbf{y}\mathbf{x}}^{\mathbf{x}}}\bar{\mathbf{w}}_{\mathbf{y}}^\mathbf{x},
\end{align}
and the inter-cluster downlink error $\boldsymbol{\epsilon}_{\text{d}_{\mathbf{y}_0}}$ is
\begin{align}
\label{interdownlinkerror} \hspace{-10pt}\frac{\vartheta_{\text{d}_{\mathbf{y}_0}}\sum_{\mathbf{x} \in \Phi_\text{p}}^{} \sqrt{\frac{{P_{\text{d}}}}{{\sum_{\mathbf{z} \in {\mathcal{C}}_\mathbf{x}}^{}\mathbb{E}\left\{\|\mathbf{v}_\text{u}^{\mathbf{z}}\|^2\right\}}}} \|\mathbf{x}+\mathbf{y}_0\|^{-\frac{\alpha}{2}} f^\mathbf{x} \sum_{\mathbf{z} \in {\mathcal{C}}_\mathbf{x}}^{}{\mathbf{v}_\text{u}^{\mathbf{z}}}}{\sqrt{\rho} \sqrt{\frac{{P_{\text{d}}}}{{\sum_{\mathbf{x} \in {\mathcal{C}}}^{}\mathbb{E}\left\{\|\mathbf{v}_\text{u}^{\mathbf{x}}\|^2\right\}}}}f^{\mathbf{o}}\|\mathbf{y}_0\|^{-\frac{\alpha}{2}}}
\end{align}
Again, the results up to here do not depend on the network topology. For the MCP-H case, we can progress as follows. The factor $\vartheta_{\text{d}_{\mathbf{y}_0}}$ is selected to minimize the distortion of the recovered
model vector ${\mathbf{w}}_{\mathbf{y}_0}^\mathbf{o}$ with respect to the ground true model vector $\frac{1}{|{\cal A}|} \sum_{\mathbf{x}\in {\mathcal{C}}}^{} \sum_{\mathbf{y}\in {\cal A}^{\mathbf{x}}}^{} {\mathbf{w}}_{\mathbf{y}}^\mathbf{x}$ from \eqref{interestimate} as
\begin{align}
\label{globaldistortion}
&\min_{\vartheta_{\text{d}_{\mathbf{y}_0}}} \mathbb{E}\left\{\left\Vert\frac{ \boldsymbol{\epsilon}_\text{u}}{|{\cal A}|} +\frac{\boldsymbol{\epsilon}_{\text{d}_{\mathbf{y}_0}}}{|{\cal A}|}\right\Vert^2\right\} \nonumber\\
&=\frac{1}{{|{\cal A}|^2}}\left(\mathbb{E}\left\{\|\boldsymbol{\epsilon}_\text{u}\|^2\right\}+\mathbb{E}\left\{\|\boldsymbol{\epsilon}_{\text{d}_{\mathbf{y}_0}}\|^2\right\}\right),
\end{align}
where due to the symmetry of the network and the error terms in \eqref{intrauplinkerror}-\eqref{intradownlinkerror} and \eqref{interuplinkerror}-\eqref{interdownlinkerror}, we have $\mathbb{E}\left\{\|\boldsymbol{\epsilon}_\text{u}\|^2\right\} =\sum_{\mathbf{x}\in {\cal C}}^{} \sum_{\mathbf{y} \in {\cal A}^\mathbf{x}}^{}\left(\vartheta_{\text{d}_{\mathbf{y}_0}}-\sigma_{\text{w},\mathbf{y}}^\mathbf{x}\right)^2+ C\frac{\vartheta_{\text{d}_{\mathbf{y}_0}}^2}{\rho}\Psi$ and $\mathbb{E}\left\{\|\boldsymbol{\epsilon}_{\text{d}_{\mathbf{y}_0}}\|^2\right\}  =\frac{{\vartheta_{\text{d}_{\mathbf{y}_0}}^{2}}\beta}{\rho|f^\mathbf{o}|^2\|\mathbf{y}_0\|^{-\alpha}} (\rho|{\cal A}|+C\Psi)$. Therefore, similar to Theorem 2, the solution of \eqref{globaldistortion} is 
\begin{align}
\label{inter_error_dowlink}
\vartheta_{\text{d}_{\mathbf{y}_0}} = \frac{\sum_{\mathbf{x} \in {\cal C}}^{}\sum_{\mathbf{y} \in {\cal A}^\mathbf{x}}^{} \sigma_{\text{w},\mathbf{y}}^\mathbf{x}}{\left(1+\frac{\beta}{|f^\mathbf{o}|^2\|\mathbf{y}_0\|^{-\alpha}}\right)\left(|{\cal A}|+\frac{C\Psi}{\rho}\right)},
\end{align}
which leads to the minimum distortion \eqref{globaldistortion} as $\frac{\sum_{\mathbf{x} \in {\cal C}}^{}\sum_{\mathbf{y} \in {\cal A}^\mathbf{x}}^{} \sigma_{\text{w},\mathbf{y}}^{\mathbf{x}\ \hspace{2pt}2}}{|{\cal A}|^2} - \frac{(\sum_{\mathbf{x} \in {\cal C}}^{}\sum_{\mathbf{y} \in {\cal A}^\mathbf{x}}^{} \sigma_{\text{w},\mathbf{y}}^\mathbf{x})^2}{\left(1+\frac{\beta}{|f^\mathbf{o}|^2\|\mathbf{y}_0\|^{-\alpha}}\right)|{\cal A}|^2\left(|{\cal A}|+\frac{C\Psi}{\rho}\right)}$.
Hence, similar to \textit{Remark 1}, the distortion of the inter-cluster aggregation in \eqref{interestimate1} has the same order $\mathcal{O}\left(\text{const}-\frac{1}{\lambda_\text{p}^2}\right)$. Here, a similar argument to \textit{Remark 2} can be made for $|{\cal A}|$.

\vspace{-4pt}
\section{Experimental Results}
The learning task over the collaborating clusters is the classification
on the standard MNIST dataset with the parameter values given in Table 1,
unless otherwise stated. The classifier model is implemented using a CNN, which consists of two $3 \times 3$ convolution layers with
ReLU activation (the first with 32 channels, the second with
64), each followed by a $2 \times 2$ max pooling; a fully connected
layer with 128 units and ReLU activation; and a final softmax
output layer. We consider both i.i.d. and non-i.i.d. distribution of dataset 
samples over the devices. For non-i.i.d. case, each device has samples of only two classes
and the number of samples at different devices is different. The performance is measured
as the learning accuracy with reference to the test dataset over global inter-cluster iteration count $t$. Each result is evaluated as the average of
10 realization.


\begin{table}
	\vspace{6pt}
	\caption {Parameter Values} 
	\vspace{-10pt}
	\begin{center}
		\resizebox{7.7cm}{!} {
			\begin{tabular}{| l | l | l | l | l | l | l | l | l | l | l | l | l | l | l |}
				
				\hline
				\hline
				{$\lambda_{\rm p}$}&{$r_0$}&{$R$}& $\mu$& $C$& $M$& $P_\text{u}$&$\left\{P_\text{d},\sigma_\text{d}^2\right\}$& $\text{th}_1$ & $\alpha$&$\tau$&$\gamma$&$T$&$B$\\ \hline
				$20$ $\text{Km}^{-2}$& $4 \ \text{m}$ &$30 \ \text{m}$&$0.01$ & 3 & 15 & 1 & $\left\{1,10\right\}$&$0.5$& 4&$6$&$2$&40&60\\ \hline	
				\hline
		\end{tabular}}
	\end{center}
	\vspace{-20pt}
\end{table}

\begin{figure*}
	\begin{subfigure}{0.33\textwidth}
		\centering
		\includegraphics[width =2in]{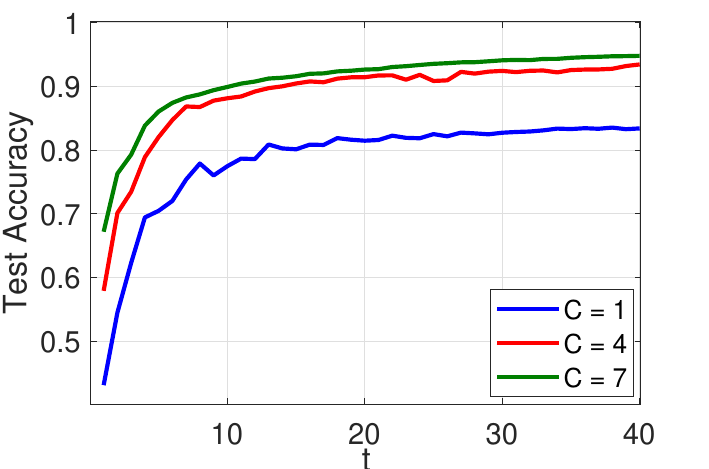} 
		\caption{}
	\end{subfigure}
	\begin{subfigure}{0.33\textwidth}
		\centering
		\includegraphics[width =2in]{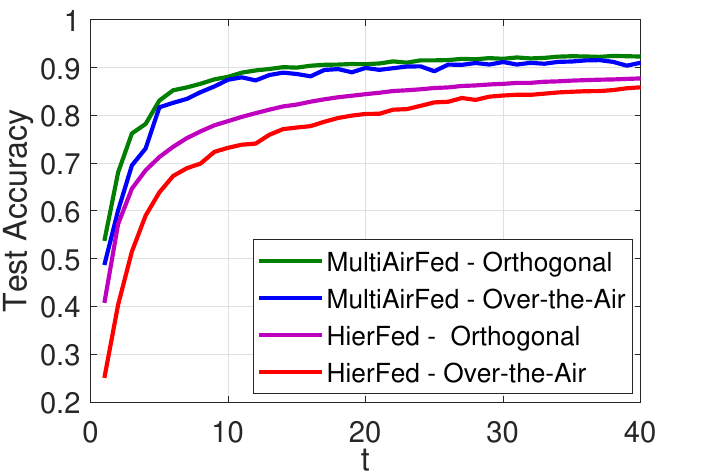} 
		\caption{}
	\end{subfigure}
	\begin{subfigure}{0.33\textwidth}
		\centering
		\includegraphics[width =2in]{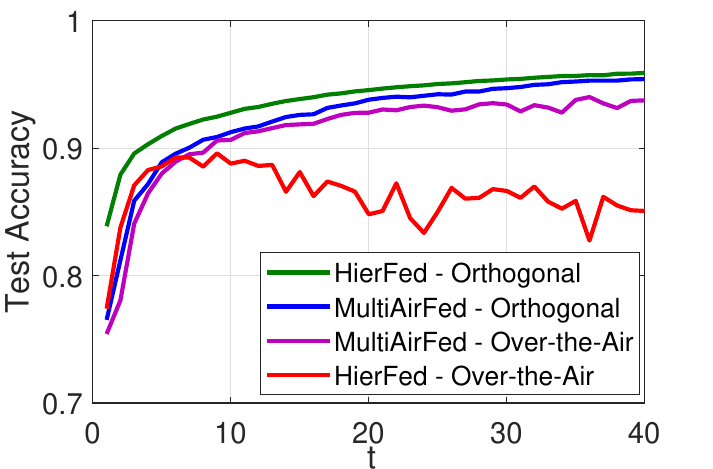} 
		\caption{}
	\end{subfigure}
	\vspace{-5pt}
	\caption{Test accuracy (a) for different $C$ values, (b) for non-i.i.d. data and  $\lambda_\text{p} = 20 \ \text{Km}^{-2}$, (c) for i.i.d. data and  $\lambda_\text{p} = 40 \ \text{Km}^{-2}$.}
	\vspace{-6pt}
\end{figure*}

In Fig. 2.a, the accuracy for different values of the number of collaborating clusters with the same task is studied in the non-i.i.d. scenario. It is observed that the multi-server case can significantly improve the accuracy compared to the single-server case. This is because the number of devices participating in the learning process increases, while the distortion decreases, according to the results of Section IV.

In Figs 2.b and c, the performance of {\fontfamily{lmtt}\selectfont MultiAirFed} is compared with the conventional hierarchical FL in \cite{letaief, bennis, tony} for both i.i.d. and non-i.i.d. distributions. We name this benchmark {\fontfamily{lmtt}\selectfont HierFed}. In {\fontfamily{lmtt}\selectfont HierFed}, the devices always upload the model parameters, and we consider $\gamma = 2$ local descent steps in each intra-cluster iteration.  
	We adopt the "over-the-air" scheme in Section IV for both  {\fontfamily{lmtt}\selectfont MultiAirFed} and {\fontfamily{lmtt}\selectfont HierFed}. Additionally, we implement both of them with "orthogonal" transmissions, which eliminates interference by assuming unlimited resources. To better evaluate the impact of interference, we consider a dense network with $\lambda_\text{p} = 40 \ \text{Km}^{-2}$ for the i.i.d. case. The results indicate that {\fontfamily{lmtt}\selectfont MultiAirFed} outperforms {\fontfamily{lmtt}\selectfont HierFed} by a substantial margin in the non-i.i.d. scenario, for both transmission schemes. This highlights the robustness of {\fontfamily{lmtt}\selectfont MultiAirFed} against both data heterogeneity and interference. In the i.i.d. scenario, although {\fontfamily{lmtt}\selectfont HierFed} outperforms {\fontfamily{lmtt}\selectfont MultiAirFed} when interference is absent, due to the multiple local steps, its performance is significantly impacted under the higher interference. In contrast, {\fontfamily{lmtt}\selectfont MultiAirFed} retains its robustness to interference. These support our reasoning in Section III.

\vspace{-10pt}
\section{Conclusions}
We proposed a new two-level federated learning algorithm that offers resilience to interference and data heterogeneity in hierarchical wireless networks. To implement the proposed algorithm independent of the network scale and with minimum resource requirements, we proposed an over-the-air aggregation scheme for the uplink and a bandwidth-limited broadcast scheme for the downlink, and derived the effect of uplink and downlink interference on the aggregations, in particular with respect to the cluster density. We showed that the proposed hierarchical FL outperforms the existing solution,
and the achieved accuracy is high, especially when the interference is high and the data is heterogeneous.
\vspace{-5pt}
\appendices
\section{Proof of Theorem 1}
The inter-cluster uplink interference power $\Psi$ is obtained as
\begin{align}
&\rho \mathbb{E}\left\{\sum_{\mathbf{x}\in \Phi_\text{p}\backslash \left\{\mathbf{o}\right\}}\sum_{\mathbf{y} \in {\cal N}^{\mathbf{x}}} {\mathds{1}}\left(|f_{\mathbf{y}\mathbf{x}}^{\mathbf{x}}|^2>\text{th}_1\right) \frac{\|\mathbf{x}+\mathbf{y}\|^{-{\alpha}}}{\|\mathbf{y}\|^{-{\alpha}}} \left|\frac{f_{{\mathbf{y}\mathbf{o}}}^\mathbf{x}}{f_{\mathbf{y}\mathbf{x}}^{\mathbf{x}}}\right|^2\right\}\nonumber\\
&= \rho\mathbb{E}\Biggl\{\sum_{\mathbf{x}\in \Phi_\text{p}}\sum_{\mathbf{y} \in {\cal N}^{\mathbf{x}}} \mathds{1}\left(|{f}_{\mathbf{y}\mathbf{x}}^{\mathbf{x}}|^2>\text{th}_1\right)\mathbb{E}\biggl\{\left.\frac{|f_{\mathbf{y}\mathbf{o}}^\mathbf{x}|^2}{|f_{\mathbf{y}\mathbf{x}}^{\mathbf{x}}|^2}\right\vert |{f}_{\mathbf{y}\mathbf{x}}^{\mathbf{x}}|^2>\text{th}_1 \nonumber\\
&\biggr\}\frac{\left(\|\mathbf{x}\|^2+\|\mathbf{y}\|^2-2\|\mathbf{x}\|\|\mathbf{y}\|\cos(\theta_\mathbf{xy})\right)^{-\frac{\alpha}{2}}}{\|\mathbf{y}\|^{-{\alpha}}} \Biggr\} \stackrel{(a)}{=} \rho\text{Ei}(\text{th}_1) \times \nonumber\\
&\mathbb{E}\Biggl\{\sum_{\mathbf{x}\in \Phi_\text{p}}\sum_{m=1}^{M} {M \choose m} e^{-\text{th}_1  m} \left(1-e^{-\text{th}_1}\right)^{M-m}m\int_{0}^{\infty}  \biggl(1+\nonumber\\
&\frac{\|\mathbf{x}\|^2}{y^2}-2\frac{\|\mathbf{x}\|}{y}\cos(\theta_{\mathbf{xy}})\biggr)^{-\frac{\alpha}{2}} \hspace{-6pt}f_{\|\mathbf{y}\|}(y)\mathrm{d}y\Biggr\} \stackrel{(b)}{=}\rho \text{Ei}(\text{th}_1)\times\nonumber\\
& Me^{-\text{th}_1} \times \lambda_\text{p} \int_{0}^{\infty}\int_{0}^{2\pi}\int_{0}^{\infty}  \left(1+\frac{x^2}{y^2}-2\frac{x}{y}\cos(\theta)\right)^{-\frac{\alpha}{2}} \nonumber\\
&f_{\|\mathbf{y}\|}(y)x \mathrm{d}y \mathrm{d}\theta \mathrm{d}x\stackrel{(c)}{=} \frac{2M \rho \lambda_\text{p} \text{Ei}(\text{th}_1)e^{-\text{th}_1}}{R^2-r_0^2}\int_{2r_0}^{\infty}\int_{0}^{2\pi}\int_{r_0}^{R}  \nonumber
\end{align}
\begin{align}
&
y\biggl(1+\frac{x^2}{y^2}-2\frac{x}{y}\cos(\theta)\biggr)^{-\frac{\alpha}{2}} x \mathrm{d}y \mathrm{d}\theta \mathrm{d}x,
\vspace{-16pt}
\end{align}
where $(a)$ is from $\mathbb{E}\biggl\{\left.\frac{|f_{{\mathbf{y}\mathbf{o}}}^\mathbf{x}|^2}{|f_{\mathbf{y}\mathbf{x}}^{\mathbf{x}}|^2}\right\vert |{f}_{\mathbf{y}\mathbf{x}}^{\mathbf{x}}|^2\hspace{-2pt}>\hspace{-2pt}\text{th}_1\biggr\} \hspace{-2pt}=\hspace{-2pt} \mathbb{E}\left\{|f_{\mathbf{y}\mathbf{o}}^\mathbf{x}|^2\right\}\mathbb{E}\bigl\{\\\left.\frac{1}{|f_{\mathbf{y}\mathbf{x}}^{\mathbf{x}}|^2}\right\vert$ $|{f}_{\mathbf{y}\mathbf{x}}^{\mathbf{x}}|^2>\text{th}_1\bigr\} \hspace{-2pt}= \hspace{-2pt}\text{Ei}(\text{th}_1)$ and $|{\cal A}^\mathbf{x}| \sim \text{Binomial}(M,\\e^{-\text{th}_1}), \forall \mathbf{x}$, $(b)$ is from the Campbell's theorem, and $(c)$ is because servers
have at least $2r_0$ distance from each other.

\end{document}